\documentclass[amssymb,prb,twocolumn,showpacs]{revtex4}
\usepackage{epsfig}
\usepackage{dcolumn}
\usepackage{amsmath}
\begin{document}
%\documentclass[12 pt,a4paper]{article} %selecciona el tipo de documento
%\usepackage[english]{babel} %selecciona el idioma
%\frenchspacing %trata los espacios despues de los puntos igual que los otros
%\usepackage{epsfig}
%\usepackage{amsmath}
%\usepackage[a4paper,dvips]{geometry}
%\geometry{textwidth=16 cm, textheight=22 cm}
%\begin{document}

%\title[Author guidelines for IOP Publishing journals in  \LaTeXe]
\title{Resonance peak shift in the photo-current of ultrahigh-mobility two-dimensional electron systems.}
%\author{J. I\~narrea and Gloria Platero}

\author{Jes\'us I\~narrea$^{1,2}$ }

\address{$^1$Escuela Polit\'ecnica
Superior,Universidad Carlos III,Leganes,Madrid,28911 ,Spain\\
$^2$Unidad Asociada al Instituto de Ciencia de Materiales, CSIC,
Cantoblanco,Madrid,28049,Spain.}

%\affiliation{Escuela Polit\'ecnica
%Superior, Universidad Carlos III, Leganes, Madrid, 28911, Spain}

%\date{\today}
%%%%%%%%%%%%%%%%%%%%%%%%%%%%%%%%%%%%%%%%%%%%%%%%%%%%%%%%%%%%%%%%%%%%%%%%%%%%%%
%\section{Abstract}
\begin{abstract}
We report on a theoretical study on the rise of strong peaks at the harmonics of the cyclotron resonance in
the irradiated magnetoresistance
in ultraclean two-dimensional electron systems. The motivation is the experimental observation  of a totally unexpected strong
 resistance peak showing  up at
the second harmonic.
We extend the radiation-driven electron orbit model (previously developed to study photocurrent oscillations and
zero resistance states) to a  ultraclean scenario that implies
 longer scattering time and longer  mean free path. Thus, when the mean free path is equivalent, in terms of energy, to  twice the cyclotron energy ($2\hbar w_{c}$),
the electron behaves as
under an  effective magnetic field twice the one really applied.
Then,  at high radiation power and/or low temperature,
a resistance spike can be observed  {\it at the second harmonic}.
For even cleaner samples the energy distance
could increase to three or four times the cyclotron energy giving rise to resistance peaks at higher harmonics (third, fourth, etc.), i.e.,
 a resonance peak shift  to lower  magnetic fields as the quality of the sample increases.  Thus, by selecting
the sample mobility one automatically would select the radiation resonance response without altering the radiation frequency.
%This effect  could be of special interest from the
%application perspective, for instance in the Terahertz band.

%These results would be of special interest from the
%application perspective, for instance in nanophotonics;
%they could lead to the design of novel ultrasensitive microwave detectors or
%to the proposal of a new generation of solar cells given
%the strong translation of radiation energy into electrical current.

\end{abstract}
%%%%%%%%%%%%%%%%%%%%%%%%%%%%%%%%%%%%%%%%%%%%%%%%%%%%%%%%%%%%%%%%%%%%%%%%%%%%%%
\maketitle
%\section{ Introduction}
Radiation-induced magnetoresistance ($R_{xx}$) oscillations (MIRO)\cite{mani1,zudov1}, show up in  high mobility
two-dimensional electron systems (2DES) when they are irradiated with microwaves (MW) at low
temperatures ($T\sim 1K$) and under low magnetic fields ($B$) perpendicular to the 2DES.
\begin{figure}
\centering \epsfxsize=3.5in \epsfysize=3.6in
\epsffile{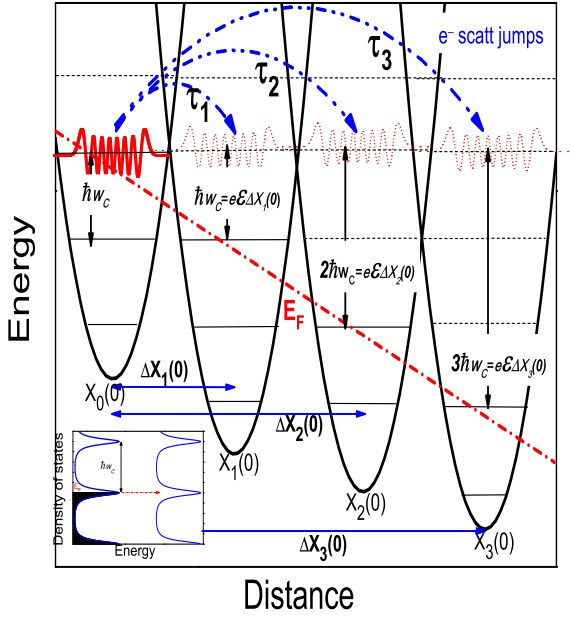}
\caption{Schematic diagram describing  elastic scattering between tilted Landau orbits (Landau states)
according to $n\hbar w_{c} = e\xi \Delta X_{n}(0)$. $\xi$ is the DC driving electric field and  $X_{n}(0)$ are the positions of the
corresponding Landau orbits. The maximum contribution to the current is obtained
when the Landau levels involved in the scattering jumps are aligned, (see inset).
}
\end{figure}
At high enough radiation  power ($P$) oscillations  maxima and
minima increase but the latter evolve into zero resistance states (ZRS)\cite{mani1,zudov1}.
Both effects  were totally unexpected when they were first obtained revealing some
type of new radiation-matter interaction assisting  electron magnetotransport\cite{ina1,ina11}.
Their discovery was considered
very important, specially in the case
of zero resistance states, because they were obtained without
quantization in the Hall resistance.
Despite the fact that over the last  years quite a few  important experimental\cite{mani2,
mani3,willett,mani4,smet,yuan,mani5,wiedmann1,wiedmann2,kons1,vk,mani6,mani61,mani62,mani7,mani71,mani100}
and theoretical efforts \cite{ina2,ina21,ina210,ina22,ina24,girvin,lei,ryzhii,rivera,ina41,ina42,ina5,ina51,ina60, ina61}
have been made on MIRO
and ZRS, their physical origin still  remains unclear and controversial.

Resonance phenomena can be found widely in nature and occur with all type of oscillations from sound to electromagnetic radiation.
They are extremely interesting in physics, from theoretical to applications perspectives, because they give rise to
an intense energy transfer between an exciting source and a driven systems.
But it turns out  definitively more intriguing  and puzzling when the resonance takes place off the
natural oscillation frequency.
This applies to one of the most challenging experimental findings\cite{yanhua,hatke} regarding MIRO and as unexpected as ZRS. It consists of a
prominent
resistance peak that  shows up at the second harmonic of the cyclotron frquency, $w \simeq 2\times w_{c}$,  ($w$ is the radiation frequency and $w_{c}$ the cyclotron frequency)
%two simultaneous effects taking place on irradiated $R_{xx}$ in ultrahigh mobility,
in irradiated $R_{xx}$\cite{yanhua,hatke} of ultrahigh mobility  2DES. This  extremely high mobility  ($\mu \geq 3\times 10^{7} cm^{2}/Vs$) along with
a low $T$  and high $P$   play an essential
role  in the appearance of this striking result.
%since they are the  only quantities that have
%significatively changed regarding previous similar experiments.
The amplitude of such a spike is very large regarding the usual MIRO, suggesting a resonance effect but off
the expected position: $w \simeq  w_{c}$.
%The second is an important collapse of $R_{xx}$ at low $B$. The latter is defined as giant (or huge) negative
%magnetoresistance (GNMR)\cite{lina1,lina2,lina3,manikri}.
% The surprising fact on the rise of the resonance spike
%to the low $T$ used in the experiments ($\sim 0.3-0.4$K),
%the
%rise of it at the second harmonic of $w_{c}$.
% when dealing with ultrahigh mobility samples.
%On the other hand, the simultaneous appearance of the off-resonance spike and
%GNMR when the sample is irradiated suggests that both in some way are closely related
%sharing, at least partially, a physical origin.
To date, there have  been presented very few
theoretical models on this topic\cite{inaspike1,bernstein}.

\begin{figure}
\centering \epsfxsize=3.3in \epsfysize=4.in
\epsffile{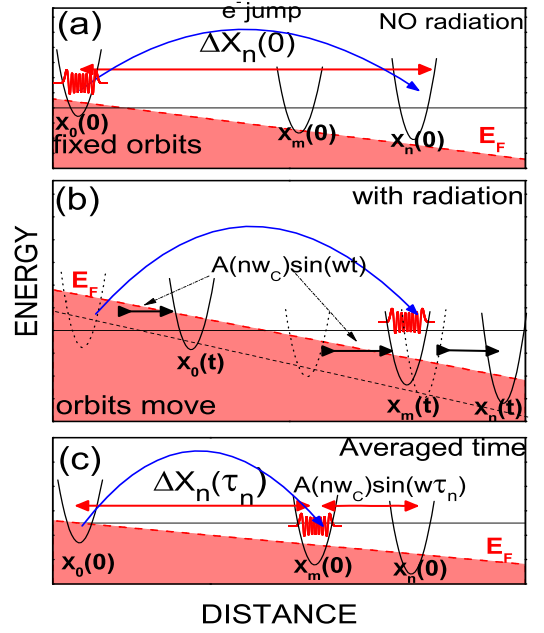}
\caption{Schematic diagrams for the emergence of a valley in MIRO in an extended scenario (distant Landau orbits). (a) Elastic scattering (charged impurities)
between Landau orbits without radiation.
(b) Elastic scattering with radiation where all Landau orbits oscillate at radiation frequency. The $X_{n}(0)$ position is occupied now by a driven-Landau
 orbit, $X_{m}(0)$, where the scattered electron lands in a time $t=\tau_{n}$. (c) With radiation but in the
steady state after time average. The final averaged distance is smaller than in the dark giving rise
to a MIRO valley. Similar reasoning can be applied for a MIRO peak.
% The diagram describes the scenario
%for a general flight time $\tau_{n}$.
%In this description the final average
%advanced distance, under  radiation and after time average, turns out to be smaller than
%in the dark. Thus, we obtain a valley in the radiation-induced oscillations.
%Similar situation but in reverse can be depicted for a peak.
}
\end{figure}

In this article, we present a theoretical analysis on this resonance peak shift
  based on the radiation-driven electron orbit model\cite{ina2,ina21}
 but adapted  to a scenario of ultra high quality samples (reduced electron scattering).
In the extension of the model we start considering that this kind of samples have
increasingly longer both mean free path and scattering time. Thus,
the scattered electron that jumps between Landau orbits (Landau states) can reach
much further, in distance and energy,  final Landau orbits (due to the DC electric field applied in the $x$ direction, see Fig. 1). For instance, orbits located at twice the
cyclotron energy ($2\hbar w_{c})$.
For this specific case the electron would behave, from the scattering standpoint, as under an {\it effective} magnetic field of double intensity than the one really applied.
Then, the spike will rise, at  low enough $T$ and high enough $P$, at the second harmonic.
For even higher mobilities we would still have longer mean free paths and then, we can predict the subsequent  rise of   $R_{xx}$ spikes at higher harmonics: $3 w_{c}=w$, $4 w_{c}=w$,   $5 w_{c}=w$, etc.,
i.e., at lower and lower $B$. Therefore and as a main result, we conclude that  by controlling the mobility of 2DES we can shift the resonance response without
altering the radiation frequency. This result could turn out  very interesting for device engineering and applications.
 For instance, irradiating a ultraclean 2D sample with Teraherzt radiation we would obtain a  resonance response in a $B$ region corresponding to MW or
even lower frequencies.

Another important result from our theoretical model is that both MIRO and  $R_{xx}$ spike would
share the same physical origin. Thus, they would stem from the interplay of the radiation-driven swinging
motion of the irradiated Landau orbits and the scattering of electrons with charged
impurities. Thus, the large $R_{xx}$ spike would corresponds to a resonance effect
 between the frequency of the Landau orbit harmonic motion and $ w_{c}$.
Obviously the former turns out to be the same as radiation frequency.
%This condition, in the model of radiation-driven
%electron orbits, corresponds to a resonance in the second harmonic: $w \simeq 2\times wc$.
%Finally, we explain the origin of the minima positions, the $1/4$ cycle phase shift.

\begin{figure}
\centering \epsfxsize=3.0in \epsfysize=3.5in
\epsffile{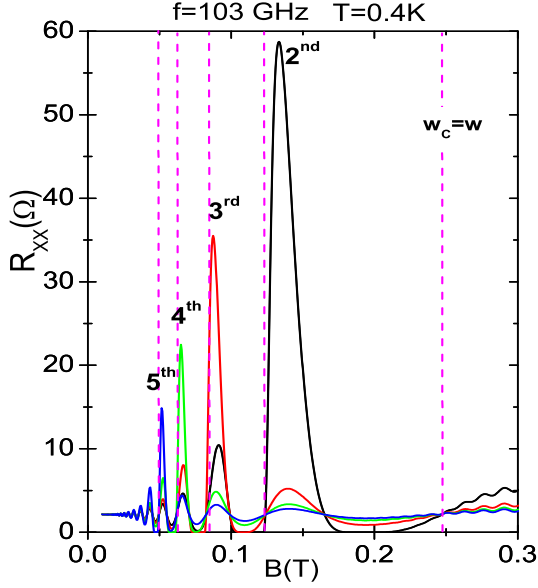}
\caption{Calculated magnetoresistance vs $B$ under radiation
of  $103  GHz$ and $T=0.4K$ for four resonance regimes: $2w_{c}=w$, $3w_{c}=w$, $4w_{c}=w$ and $5w_{c}=w$. Vertical dashed lines  mark the harmonic positions.
Spikes rise up from the second to fifth harmonic. The photo-excited oscillations positions remain constant showing the
1/4 cycle phase shift independently of
the resonance peak displacement.
}
\end{figure}

\begin{figure}
\centering\epsfxsize=3.in \epsfysize=5.5in
\epsffile{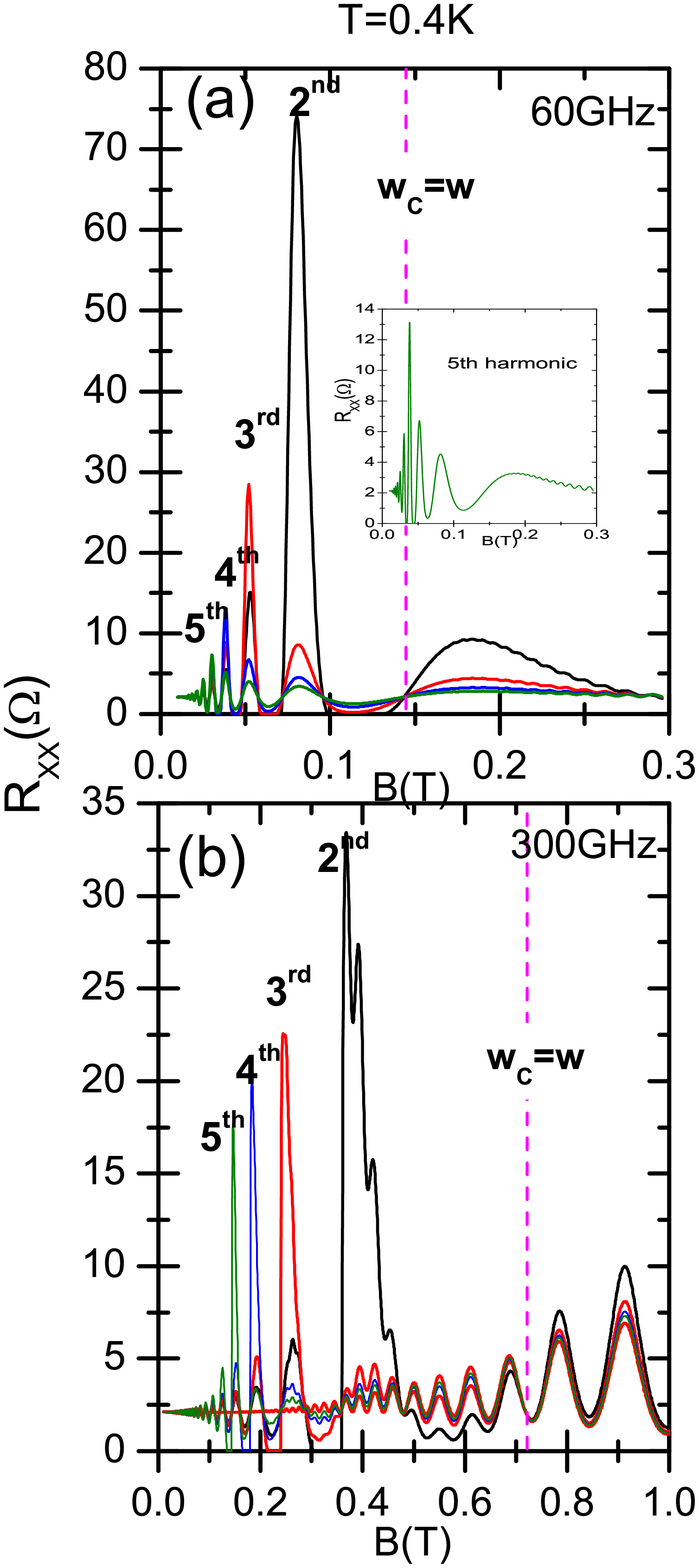}
\caption{ Same as in  Fig.3.  We exhibit the same  four off-resonance spikes for $60 GHz$ in panel (a) and for
$300 GHz$ in panel (b). The vertical dashed line for both panels corresponds to the main resonance frequency.
The inset in panel (a) is a zoom-in of the fifth harmonic. }
\end{figure}

%\section{ Theoretical Model}
The {\it radiation-driven electron orbits model}, was devised  by the authors to address
two physical effects triggered by radiation: MIRO and ZRS in high  mobility  2DES. According to this model, under radiation the
 Landau orbits, spatially and  harmonically  oscillate with the radiation frequency. As a result, the scattering process  of
electrons with charged impurities turns out
to be dramatically   altered.
This is reflected in magnetotransport and in turn in $R_{xx}$
giving rise to the well-known MIRO
 and  ZRS\cite{ina2,ina21,ina30,kerner,park}.
Following the model, we use a semiclassical Botzmann theory
to calculate the longitudinal conductivity $\sigma_{xx}$\cite{ando,ridley,miura}:
\begin{equation}
\sigma_{xx}=2e^{2} \int_{0}^{\infty} dE \rho_{i}(E) (\Delta X)^{2}W_{I}\left( -\frac{df(E)}{dE}  \right)
\end{equation}
being $E$ the energy and $\rho_{i}(E)$ the density of
initial Landau states.  The expression for $\Delta X$ is likewise obtained from the  model\cite{inahole,inarashba}:
\begin{equation}
\Delta X= \Delta X_{1}(0)-A(w_{c})  \sin \left(w \tau_{1}\right)
\end{equation}
%e^{-\pi\frac{\gamma}{w_{c}}}
where $\Delta X_{1}(0)$ is the distance between the guiding centers of the final and initial
Landau orbits in the dark and $\tau_{1}=2\pi/w_{c}$  is the  {\it flight time}: the time it takes the
scattered electron between Landau orbits.  It was previously proposed, in a semiclassical
approach\cite{inahole,nazareno}, that
during the scattering jump
 electrons in their orbits would complete, on average, an
integer number of cyclotron orbits,
which implies that $\tau_{n}=n \times T_{c}=n \times \frac{2 \pi}{w_{c}}$, $T_{c}$ being the cyclotron time.
%The reason is that during the scattering jump between Landau orbits,
%the carriers in their orbits complete one full loop in the flight time $\tau_$.
%This implies that  = Tc. 
Therefore, the carrier involved in
the scattering ends up in the same relative position inside the
final orbit as the one it started from in the initial one. The
reason for this is that the dynamics of the orbits (Landau
states) is governed on average by the position of the centre
of the orbit irrespective of the carrier position inside the orbit
when the scattering takes place. Then, on average, both the
initial and final semiclassical positions are identical in their
respective orbits
%This stems from the point that in high mobility samples,
%scattering between
%Landau orbits  is governed   by the position of the orbit guiding centre.
%Then the relative electron position in the orbit has to keep constant before and after
%the scattering jump\cite{nazareno}.
In the  radiation-driven electron orbit model\cite{inahole,inarashba} $n=1$ and  $\tau_{1}=T_{c}$
corresponding to one cyclotron orbit during the jump.
% And then to keep constant this distance, the carrier position inside the orbit
%has to be the same at the beginning and end of scattering\cite{nazareno}.
$A(w_{c})$ is the amplitude of
the spatial  oscillations of the driven orbits in the $x$ direction:  $A(w_{c})=\frac{e E_{0}\sin wt}{m^{*}\sqrt{w^{2}(w_{c}-w)^{2}+\gamma^{4}}}$,
$E_{0}$ being the radiation electric field and $\gamma$ is a damping parameter related to
the interaction of the electrons in the driven Landau orbits with the lattice ions.
$W_{I}$ is the scattering rate of
electrons with charged impurities that according to the Fermi's golden rule: $W_{I}=\frac{2\pi}{\hbar}|<\phi_{f}|V_{s}|\phi_{i}>|^{2}\delta(E_{i}-E_{f})$,
where $\phi_{i}$ and $\phi_{f}$ are the wave functions  corresponding to the initial and final Landau states respectively and
 $V_{s}$ is the scattering potential for charged impurities\cite{ando}. The expressions of the initial and final energies
 are: $E_{i}=\hbar w_{c}(i+1/2)$ and $E_{f}=\hbar w_{c}(f+1/2)-\Delta_{DC}$, where $i$ and
$f$ are integers and $\Delta_{DC}=e\xi \Delta X(0)$, $\xi$ is the DC driving electric field
 in the x direction and
responsible of the current along that direction (see Fig. 1).
To obtain $R_{xx}$ we use the common tensorial relation
$R_{xx}=\frac{\sigma_{xx}}{\sigma_{xx}^{2}+\sigma_{xy}^{2}}
\simeq\frac{\sigma_{xx}}{\sigma_{xy}^{2}}$, where
$\sigma_{xy}\simeq\frac{n_{i} e}{B}$, $n_{i}$ being the electrons density, e the electron charge and $\sigma_{xx}\ll\sigma_{xy}$.

According to $W_{I}$ the largest contributions to the conductivity in
the presence of the field $\xi$ occurs  when $E_{i}=E_{f} \Rightarrow e\xi \Delta X_{n}(0)\simeq n \times \hbar w_{c}$,
implying that the Landau level indexes, $f$ and $i$  are related by $f=i+n$, $n$ being a positive integer.
Or in other words, when the Landau levels are aligned (see inset in Fig.1).
The   $n=1$ scenario labelled with $\tau_{1}$ in Fig.1,  implies that
 $e\xi \Delta X_{1}(0)= \hbar w_{c}$ and it would correspond to ordinary MIRO.
In a general  extension of the model we can include other scattering processes
that are, likewise, likely to happen according to $W_{I}$ (their Landau levels are
aligned too).
Theses  processes are labelled in Fig.1,  with  $\tau_{2}$ and $\tau_{3}$  (flight times of theses processes) and correspond to
energy differences (in reference to the Fermi energy) of $e\xi \Delta X_{2}(0)= 2\hbar w_{c}$ and
$e\xi \Delta X_{3}(0)= 3\hbar w_{c}$ respectively (see Fig.1).  Yet, the long distance
between the Landau orbits involved in the scattering (longer than the mean free path in ordinary samples) prevent them from happening;
the corresponding probability is very small. Accordingly, and in light of the uncertainty principle\cite{cohen},  the minimum values of
those flight times can be obtained: $\tau_{2}=2\pi/2w_{c}$ and $\tau_{3}=2\pi/3w_{c}$ respectively. Thus, we obtain increasingly shorter
flight times in increasingly longer scattering times. And the flight times get smaller not in a continuous way but by integer values of $w_{c}$ in the denominator.

Interestingly enough and according to the above, the process labelled with $\tau_{2}$ would be mainly described by scattering
quantities such as the distance
between Landau orbits, $\Delta X_{2}(0)$ and the corresponding flight time, being both given by:
\begin{eqnarray}
e\xi \Delta X_{2}(0)&=& \hbar w_{c}^{\star}\\
\tau_{2}&=&\frac{2\pi}{w_{c}^{\star}}
\end{eqnarray}
where $ \hbar w_{c}^{\star}=\hbar \frac{eB^{\star}}{m^{\ast}}=\hbar \frac{e(2B)}{m^{\ast}}=2\hbar w_{c}$.
Then, from the scattering point of view,  we would obtain the same results as if the electron
were under an effective twice as high magnetic field ($B^{\ast}$) than  the one really applied ($B$).
Similar approach can be applied to the $\tau_{3}$ and further scenarios. Thus, for  $\tau_{3}$ we
would have and effective magnetic field $B^{\star\star}=3B$ and $w_{c}^{\star\star}=3w_{c}$.
Therefore in our model the increasing quality of the sample makes vary the main scattering quantities
in steps, i.e., in integer values of $w_{c}$.
%and $\tau_{3}$ we would obtain
%similar scattering  results to the ones of a $\tau_{1}$  process, (energy difference of $\hbar w_{c}$), but under double or triple $B$ respectively;
%as if the electron were scattered under an effective $B$ twice or three times larger than the one really applied.
The above discussion is essential for the model and  would be at the heart
of the experimental results as we explain below.
%In other words,
%starting from the $\tau_{1}$ scenario, by increasing $B$ we would end up reaching scattering situations similar
%to the $\tau_{2}$ or $\tau_{3}$ scenarios, when $B$ were twice or three times bigger respectively.
Now, applying the theory of radiation driven electron orbit model\cite{inarashba} to these new scenarios, (from $\tau_{1}$  to $\tau_{2}$
and $\tau_{3}$ etc.), we obtain a general expression for $\Delta X$,
\begin{widetext}
\begin{equation}
\Delta X= [\Delta X_{1}(0)-A(w_{c})  \sin \left(w \tau_{1}\right)]+[\Delta X_{2}(0)-A(w_{c}^{\star})  \sin \left(w \tau_{2}\right)]+
[\Delta X_{3}(0)-A(w_{c}^{\star\star})  \sin \left(w \tau_{3}\right)]+.....
\end{equation}
\end{widetext}
where $A(w_{c}^{\star})=\frac{e E_{0}\sin wt}{m^{*}\sqrt{w^{2}(w_{c}^{\star}-w)^{2}+\gamma^{4}}}$ and
$A(w_{c}^{\star\star})=\frac{e E_{0}\sin wt}{m^{*}\sqrt{w^{2}(w_{c}^{\star\star}-w)^{2}+\gamma^{4}}}$. And accordingly,
we could obtain the resonance peak in different $B$-positions depending on what is the predominant term over the rest.
In the general  expression above, we have extended the basic idea of our model  that when the radiation is on,
the Landau orbits oscillate (driven by radiation) altering the electron scattering. And for different
flight times (depending of the scenario)   $\tau_{n}=2\pi/n w_{c}$, the scattered electron
will be {\it landing} in a different final Landau orbit and different in turn from the dark giving rise to a distance shift  in the scattering jump.
After averaging out, the shift  is given by $A(nw_{c})  \sin \left(w \tau_{n}\right)$  (see Fig. 2). This shift can be positive or negative (or zero)
and is  finally reflected in $\sigma_{xx}$ and $R_{xx}$ in the form of peaks and valley respectively, i.e., photo-excited $R_{xx}$ oscillations or MIRO.

\begin{figure}
\centering\epsfxsize=3.in \epsfysize=3.0in
\epsffile{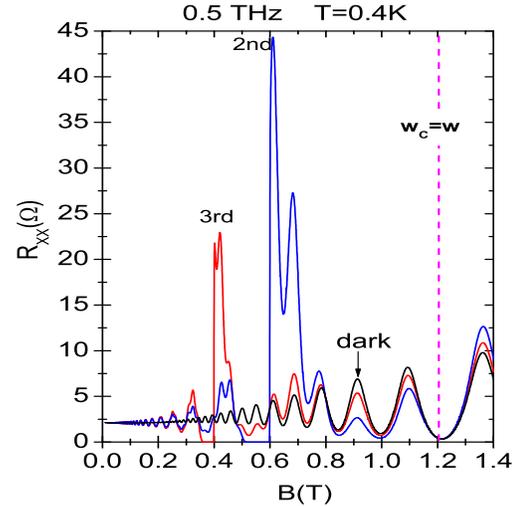}
\caption{Same as in  Fig.3., for a frequency of $0.5THz$. We exhibit two off-resonance spikes for
the second and third harmonic positions. The vertical dashed line  corresponds to the main resonance frequency.}
\end{figure}

As we said above, in an ordinary MIRO experiment only the expression of
the first bracket in the right part of  the latter equation, (the  $[\Delta X_{1}(0)-A(w_{c})  \sin \left(w \tau_{1}\right)] $   term), would significantly contribute to $\Delta X$.  In this regime
the shorter mean free path prevent from reaching further Landau orbits making negligible the
contributions of the other terms.
Nevertheless, when it comes to ultrahigh  mobility samples we will have longer mean free paths and
  scattering times, and much further final Landau orbits can be accessible via scattering.
Thus, by increasing mobility  we would end up having first the $[\Delta X_{2}(0)-A(2w_{c})  \sin \left(w \tau_{2}\right)]$ term as predominant
where the resonance peak would rise at the second harmonic. In a next step increasing further mobility even more distant
Landau orbits would be accessible and thus, we would obtain the
third term, ($[\Delta X_{3}(0)-A(3w_{c})  \sin \left(w \tau_{3}\right)]$),  as predominant and  the resonance at the
third harmonic, etc. For instance, what it is obtained in the off-resonance experiments\cite{yanhua,hatke} would be
based on the second term and the expression of the average advanced distance would be: $\Delta X\simeq [\Delta X_{2}(0)-A(2w_{c})  \sin \left(w \tau_{2}\right)]$. But still
the flight time $\tau_{2}$ needs to be adapted to a ultra-clean scenario, i.e., it has to be increased as the scattering time increases too.
As we said above in our model, this increment has to be made in multiples of $T_{c}$
corresponding to the effective magnetic field.
This implies that now  the new $\tau_{2}= 2\times 2\pi/2 w_{c}$. Substituting in $\Delta X$, we
finally obtain: $\Delta X\simeq \left[\Delta X_{2}(0)-A(2w_{c})  \sin \left(2\pi\frac{w}{w_{c}}\right)\right]$.
The latter is a remarkable result because it can be generalized to higher order terms where,
as the magnetoresistance resonance peak shifts to lower $B$, the photo-excited oscillations (MIRO) would
remain at the same $B$-position. This is in agreement with experiments\cite{yanhua,hatke}.

In Fig. 3 we exhibit the calculated magnetoresistance vs $B$ under radiation of $103 GHz$ and $T=0.4K$.
We present four curves each one corresponding to a different resonance regime: $2w_{c}=w$, $3w_{c}=w$, $4w_{c}=w$ and $5w_{c}=w$.
Spikes are obtained from the second to fifth harmonic. The harmonic positions are given by the dashed
vertical lines (including the main resonance). And as explained above, the oscillations positions remain constant showing the 1/4 phase shift independently of the
resonance peak displacement that moves to lower $B$ for each harmonic. In Fig.4 we present similar results to Fig.3 but for
two distant frequencies of the microwave band to prove that the off-resonance spikes are a generic  feature
of MIRO. However they can only be observed clearly with ultraclean 2DES, cleaner that
the ordinary conditions to observe MIRO and ZRS. We exhibit the four off-resonance spikes for $60 GHz$ in panel (a) and for
$300 GHz$ in panel (b). The vertical dashed line for both panels corresponds to the main resonance frequency.
In Fig. 5 we exhibit calculated results corresponding to the terahertz band for a radiation frequency of
$0.5 THz$. Thus, we present  the same as in Fig.3 but for only two irradiated curves corresponding
to the $2w_{c}=w$ and  $3w_{c}=w$ resonance regimes.  Thus, on the one hand,  we want to demonstrate that this effect
can show up at higher frequencies than the microwave band proving that it can be considered as a universal effect.
And, on the other hand, from the application standpoint it might be interesting to stress that for a terahertz
frequency we can obtained the resonance response clearly inside the microwave  range just by increasing the mobility of the sample.

%\section{Conclusions}
In summary, we have presented a theoretical approach  on the off-resonance spikes generation  in irradiated magnetoresistance in ultraclean 2DES.
We have explained the experiments where the spike at the second harmonic is obtained and predict the
appearance of subsequent spikes at higher harmonics for higher mobilities.
We have explained this striking effect from the perspective of the radiation-driven
electron orbits model based on the idea that this kind of samples have
a longer mean free path. Thus,   when, in terms of energy, this distance is  twice the cyclotron energy,
the electron behaves as
under an  effective $B$ double than the one really applied. Then
a resistance spike {\it at the second harmonic} can be observed.

%time it takes the electron to jump, triggered by charged impurity scattering,  from one Landau orbit to
%another is the core variable that rules both effects. We have  predicted the rise of higher harmonics at
%higher mobilities giving numerical estimates in terms of sample mobility.
%We also explain the origin of the minima positions and their $1/4$ cycle phase shift.

%\section{Acknowledgments}

This work is supported by the MINECO (Spain) un-
der grant MAT2017-86717-P and ITN Grant 234970 (EU).
GRUPO DE MATEMATICAS APLICADAS A LA MATERIA CONDENSADA, (UC3M),
Unidad Asociada al CSIC.

%\section{References}

\end{document}